\documentclass[11pt,a4paper,twocolumn]{article}

\usepackage[a4paper,margin=1.8cm,columnsep=0.65cm]{geometry}
\usepackage{graphicx}
\usepackage{natbib}
\usepackage{amssymb,amsmath}
\bibpunct{(}{)}{;}{a}{}{,}
\usepackage{booktabs}
\usepackage[breaklinks=true,colorlinks=true,linkcolor=blue,citecolor=blue,urlcolor=blue]{hyperref}

\title{Relativistic Time Scales and Transformations in the Solar System}
\author{%
  Hong-Bo Jin$^{1,*}$,
  Jinsong Ping$^{1}$,
  Min Liu$^{2}$,
  and Mingyuan Wang$^{1}$
}

\renewenvironment{abstract}{%
  \small
  \quotation
}{%
  \endquotation
}

\begin{document}

\twocolumn[%
  \maketitle
  \vspace{0.6em}
  \begin{abstract}
Each solar-system observable is characterised by celestial reference system (CRS) coordinate time, proper time on its world line, and the transformation between them. Ephemerides and Deep Space Network (DSN) tracking use the International Astronomical Union (IAU) barycentric and body-centric hierarchy, now extended to cislunar and Mars work.
The IERS Conventions, Moyer radiometric models, and recent lunar-time papers distribute metric, scale, and tracking formulae across separate manuals. Merged Chang'e- or Tianwen-class data can acquire microsecond-level range and Doppler biases unless proper time $\tau$ is mapped consistently to barycentric and body-centric coordinate times.
We present a unified 1PN documentation chain: tabulated harmonic Christoffel symbols through $\mathcal{O}(c^{-4})$, the barycentric--geocentric--terrestrial coordinate-time sequence, Fermi normal coordinates, null-geodesic observables, and a 1PN two-way range-rate expansion, applied in parallel to Mars (MCRS/MCG) and lunar (LCRS/TCL) body-centric systems.
The chain yields a Mars areoid--geoid metric clock-rate difference of $\sim$48~$\mu$s\,day$^{-1}$ and lunar selenoid--geoid rates of $\sim$57.4--58.7~$\mu$s\,day$^{-1}$ consistent with published nested coefficients. Mars-range Shapiro-rate terms reach $10^{-12}$--$10^{-13}$. Multi-CRS consistency relies on documented transformation chains rather than a single master clock.
  \end{abstract}
  \vspace{0.4em}
  \noindent\textbf{Keywords:} relativistic processes --- reference systems --- Moon --- Mars --- methods: analytical --- planets and satellites: general
  \vspace{0.6em}
]
\footnotetext[1]{\small National Astronomical Observatories, Chinese Academy of Sciences, Beijing 100101, China; {\it hbjin@bao.ac.cn, wangmy@nao.cas.cn}}
\footnotetext[2]{\small Beijing Orient Institute of Metrology and Test, Beijing 100094, China}
\begingroup
\renewcommand{\thefootnote}{*}
\footnotetext{\small Corresponding author. E-mail: \href{mailto:hbjin@bao.ac.cn}{hbjin@bao.ac.cn}}
\endgroup

\section{Introduction}
\label{sec:intro}

Solar-system time measurement is not a single time label but a \emph{time-scale specification}. Each observable is characterised by the celestial reference system (CRS) that defines coordinate time, the world line that carries proper time $\tau$, and the transformation between them. Coordinated Universal Time (UTC) and Terrestrial Time (TT) enter only as boundary conditions at ground stations. Ephemerides, radiometric measurement models, and frequency-transfer data are formulated in the Barycentric Celestial Reference System (BCRS) or in body-centric celestial reference systems and reduced through documented time transformations \cite{soffel2003iau,kaplan2005iau,petit2010iers}. Reviews by Ashby \cite{ashby2003relativity}, Brumberg \cite{brumberg1995essential}, and the International Earth Rotation and Reference Systems Service (IERS)/Moyer formalism \cite{moyer2003formalism,petit2010iers} remain the standard entry points for geocentric work.

Interest in timekeeping beyond Earth has accelerated sharply since 2024. The International Astronomical Union (IAU) General Assembly adopted Resolution~II defining the Lunar Celestial Reference System (LCRS) and Lunar Coordinate Time (TCL). Resolution~III invites international agreement on a coordinated lunar time standard \cite{iau2024lcrs,iau2024ltime}. In parallel, the United States Office of Science and Technology Policy directed NASA to develop Coordinated Lunar Time (LTC) for cislunar operations by late 2026 \cite{whitehouse2024celestial}. The Bureau International des Poids et Mesures (BIPM) convened lunar-time working groups that compare three reference-timescale options for the Moon \cite{bourgoin2026lunar}. On the relativistic side, Kopeikin \& Kaplan \cite{kopeikin2024lunar}, Ashby \& Patla \cite{ashby2024relativistic}, and Turyshev et al.\ \cite{turyshev2025lunar} derived TCL and TT surface clock-rate transformations from the IAU 2000 framework. Bourgoin et al.\ \cite{bourgoin2026lunar} consolidated these results into operational options for a lunar reference timescale (TL) steered to TCL. In China, Liu et al.\ \cite{liu2025convert,liu2025rule} gave nested Earth--Moon time transformations and lunar timekeeping dissemination rules. Related lunar navigation work now assumes microsecond-level CRS documentation. Examples include orbit-based clock alignment \cite{yang2026lunar} and Queqiao-type relay constellations that combine lunar communication, navigation, and time transfer. Mars and other bodies appear in the same policy documents. A comparable Mars Celestial Reference System (MCRS) and Mars Coordinate Time (MCG) operational chain remains largely prospective.

The IERS Conventions, Moyer formulation, and recent lunar-time papers treat metric coefficients, coordinate scales, and radiometric models in separate manuals \cite{petit2010iers,moyer2003formalism,liu2025convert,kopeikin2024lunar}. For merged Chang'e- or Tianwen-class tracking, mission teams need a single 1PN map from the harmonic line element through IAU times to two-way range and Doppler partials, with gravitomagnetic terms and explicit $\mathcal{O}(c^{-4})$ corrections kept in one place.

This paper is a unified \emph{documentation chain}, not a replacement for the IAU resolutions or for nested Earth--Moon transforms already given by Liu et al.\ \cite{liu2025convert,liu2025rule}, Kopeikin \& Kaplan \cite{kopeikin2024lunar}, and Ashby \& Patla \cite{ashby2024relativistic}. Relative to those works and to earlier Mars $\tau$--TCG studies in RAA \cite{pan2014mars,xu2016mars}, our incremental contributions are:
\begin{enumerate}
\item a tabulated harmonic Christoffel set through $\mathcal{O}(c^{-4})$ with $V_i$, linked to the redshift and null-geodesic routines used in IERS-class software;
\item an explicit 1PN expansion of the two-way geometric range rate $\dot{\rho}$, where $\rho$ is the instantaneous station--spacecraft range (Section~\ref{sec:light}), including Shapiro-rate and Sagnac-type terms at Mars-range sensitivity;
\item a parallel MCRS/MCG prescription alongside LCRS/TCL within the same body-centric template, with areoid--geoid and selenoid--geoid rate estimates cross-checked against published lunar coefficients.
\end{enumerate}
The paper is organised as follows. Section~\ref{sec:terrestrial} fixes the time-scale vocabulary used as boundary data at ground stations. Sections~\ref{sec:kinematics}--\ref{sec:light} develop the formal chain from the clock hypothesis and PPN metric through IAU coordinate times, Fermi normal coordinates, and two-way Doppler observables. Section~\ref{sec:body-crs} applies the chain to Mars and the Moon. Section~\ref{sec:operations} states operational implementation, and Section~\ref{sec:conclusion} summarises the results.

\section{Time scales in solar-system metrology}
\label{sec:terrestrial}

Solar-system metrology uses a hierarchy of time labels from laboratory atomic scales through barycentric and body-centric coordinate times. The SI second is defined by the unperturbed caesium-133 hyperfine transition, with primary realisations referred to the geoid \cite{bipm2019si}. International Atomic Time (TAI) is a continuous atomic scale. Coordinated Universal Time satisfies $t_{\mathrm{UTC}}=t_{\mathrm{TAI}}-\Delta\mathrm{LS}$ with integer leap seconds so that $|t_{\mathrm{UTC}}-t_{\mathrm{UT1}}|\le 0.9$~s \cite{nelson2001leap}. At ground stations, UTC is the usual operational time tag for tracking data. Terrestrial Time (TT) and Geocentric Coordinate Time (TCG) support laboratory comparisons and geocentric work on Earth \cite{soffel2003iau,kaplan2005iau,standish1998time}. Ephemerides and radiometric measurement models are formulated in barycentric coordinate time (TCB) or Barycentric Dynamical Time (TDB) as a linearly scaled TCB. Table~\ref{tab:timescales} lists these scales together with the provisional lunar and Mars labels developed in Section~\ref{sec:body-crs}. Each data product identifies which label applies and through which time transformation it is tied to $\tau$ on a specified world line.

\begin{table}[t]
\centering
\caption{Time Scales in Solar-System Metrology.}
\label{tab:timescales}
\small
\begin{tabular}{@{}p{0.18\columnwidth}p{0.62\columnwidth}@{}}
\toprule
Scale & Role \\
\midrule
TAI/UTC & Terrestrial atomic time scale; UTC linked to UT1 \\
TT      & Geocentric proper-time surrogate for ephemerides \\
TCG     & Geocentric Celestial Reference System (GCRS) coordinate time \\
TCL     & Lunar Celestial Reference System (LCRS) coordinate time (provisional) \\
LTC     & Proposed operational lunar scale steered to TCL \\
MCG     & Mars Celestial Reference System (MCRS) coordinate time (provisional) \\
TDB     & Barycentric Dynamical Time; linearly scaled TCB (no average drift vs.\ TT at geocenter) \\
TCB     & Barycentric Celestial Reference System (BCRS) coordinate time \\
$\tau$  & Invariant proper time on a world line \\
\bottomrule
\end{tabular}
\end{table}

\section{Spacetime kinematics and the clock hypothesis}
\label{sec:kinematics}

Along a timelike world line $x^{\mu}(\lambda)$, proper time is defined by $\mathrm{d}\tau=c^{-1}\sqrt{-\mathrm{d}s^{2}}$ with $\mathrm{d}s^{2}=-c^{2}\mathrm{d}t^{2}+\mathrm{d}\vec{x}^{\,2}$ in Minkowski space \cite{will1993theory,ashby2003relativity}. In a local inertial frame this yields the time-dilation law
\begin{equation}
  \mathrm{d}\tau = \sqrt{1-\frac{v^{2}}{c^{2}}}\;\mathrm{d}t.
  \label{eq:sr-tau}
\end{equation}
The \emph{clock hypothesis} identifies $\tau$ with the reading of an ideal standard transported along its world line. In a Lorentzian manifold $(\mathcal{M},g_{\mu\nu})$ with signature $(-,+,+,+)$,
\begin{equation}
  \mathrm{d}\tau^{2} = -\frac{1}{c^{2}}\,g_{\mu\nu}\,\mathrm{d}x^{\mu}\mathrm{d}x^{\nu}.
  \label{eq:curved-tau}
\end{equation}
Equation~(\ref{eq:curved-tau}) reduces to (\ref{eq:sr-tau}) in a local inertial frame. Remote clock comparison requires a specified procedure (parallel transport, Fermi--Walker tetrad, or coordinate-time integration; Section~\ref{sec:fermi}); there is no coordinate-independent distant simultaneity \cite{damour2002coordinate}.

\section{Post-Newtonian spacetime and PPN parameters}
\label{sec:ppn}

\subsection{PPN Metric in Harmonic Coordinates}

Solar-system ephemerides employ a post-Newtonian expansion of the Einstein equations. Throughout Section~\ref{sec:ppn} and the radiometric treatment below, 1PN accuracy means retaining metric and clock-rate terms through $\mathcal{O}(c^{-2})$, with Christoffel symbols and selected observables extended to $\mathcal{O}(c^{-4})$ where noted. In harmonic coordinates $(t,\vec{x})$ the standard PPN form \cite{will1993theory,soffel2003iau} reads, to $\mathcal{O}(c^{-4})$,
\begin{align}
  \mathrm{d}s^{2} &= A\,c^{2}\mathrm{d}t^{2}
  + \frac{2(1+\gamma)}{c^{2}}\sum_{i} V_{i}\,\mathrm{d}t\,\mathrm{d}x^{i}
  \notag\\
  &\quad - \left[1+\frac{2\gamma U}{c^{2}}\right]\mathrm{d}\vec{x}^{\,2},
  \label{eq:ppn}\\
  A &= 1-\frac{2U}{c^{2}}+\frac{2\beta U^{2}}{c^{4}}
  -\frac{(1+2\gamma)}{c^{4}}\,\mathcal{K},
  \notag
\end{align}
with $\mathcal{K}=\sum_{i}\int G\rho(\vec{x}')\,\mathrm{d}^{3}x'/|\vec{x}-\vec{x}'|$, $U=\sum_{A} G M_{A}/r_{A}$ the Newtonian potential, $V_{i}$ the gravitomagnetic potential, and $(\gamma,\beta)$ are PPN parameters ($\gamma=\beta=1$ in general relativity). Present bounds from spacecraft tracking and lunar laser ranging give $|\gamma-1|\lesssim 10^{-5}$ \cite{will2014confrontation}. The isotropic 1PN metric used in many ephemeris manuals,
\begin{align}
  \mathrm{d}s^{2} &=
  \left(1-\frac{2w}{c^{2}}-\frac{v^{2}}{c^{2}}\right)c^{2}\mathrm{d}t^{2}
  \notag\\
  &\quad - \left(1+\frac{2w}{c^{2}}\right)\mathrm{d}\vec{x}^{\,2}
  + \mathcal{O}(c^{-4}),
  \label{eq:pn-metric}
\end{align}
is obtained from (\ref{eq:ppn}) after a generating function to isotropic coordinates with $w$ the total $n$-body potential in the selected CRS \cite{soffel2003iau}.

\subsection{Christoffel Symbols and Geodesics}

For any vector $X^{\alpha}$ on a timelike world line with four-velocity $u^{\alpha}=\mathrm{d}x^{\alpha}/\mathrm{d}\tau$, the \emph{covariant derivative along the world line} is
\begin{equation}
  \frac{\mathrm{D}X^{\alpha}}{\mathrm{D}\tau}
  \equiv u^{\beta}\nabla_{\beta}X^{\alpha}
  = \frac{\mathrm{d}X^{\alpha}}{\mathrm{d}\tau}
  + \Gamma^{\alpha}_{\beta\gamma}\,u^{\beta}X^{\gamma},
  \label{eq:covariant-dtau}
\end{equation}
where $\Gamma^{\alpha}_{\beta\gamma}$ are the Christoffel symbols of the embedding CRS metric ((\ref{eq:G000})--(\ref{eq:Gijk})). The four-acceleration is $a^{\alpha}\equiv\mathrm{D}u^{\alpha}/\mathrm{D}\tau$; it vanishes on a freely falling geodesic, in which case (\ref{eq:covariant-dtau}) reduces to \emph{parallel transport}.

Freely falling test particles follow $\mathrm{D}u^{\mu}/\mathrm{D}\tau=0$, i.e.
\begin{equation}
  \frac{\mathrm{d}^{2}x^{\mu}}{\mathrm{d}\tau^{2}}
  + \Gamma^{\mu}_{\alpha\beta}\,u^{\alpha}u^{\beta} = 0,
  \label{eq:geodesic}
\end{equation}
with $\Gamma^{\mu}_{\alpha\beta}$ built from $g_{\mu\nu}$. At 1PN order the spatial Christoffel symbols produce the $\mathcal{O}(v/c)$ gravitational light deflection and the gravitomagnetic Coriolis-type accelerations proportional to $(1+\gamma)$. Clocks on bound orbits are \emph{not} on timelike geodesics unless they are truly freely falling; spacecraft thrust and planetary rotation introduce non-geodesic terms that are added to (\ref{eq:geodesic}) when modelling $\tau$.

\subsection{Coordinate-Time Rate Along a World Line}

Along a world line with coordinate velocity $\vec{v}=\mathrm{d}\vec{x}/\mathrm{d}t$,
\begin{equation}
  \frac{\mathrm{d}\tau}{\mathrm{d}t}
  = \left(1+\frac{w}{c^{2}}-\frac{v^{2}}{2c^{2}}\right)
  + \mathcal{O}(c^{-4}),
  \label{eq:gr-clock}
\end{equation}
the 1PN specialization of (\ref{eq:curved-tau}). The redshift ratio between two clocks instantaneously at rest at potentials $w_{1,2}$ is
\begin{equation}
  \frac{\mathrm{d}\tau_{2}}{\mathrm{d}\tau_{1}}
  = \frac{1+w_{2}/c^{2}}{1+w_{1}/c^{2}}
  \approx 1+\frac{w_{2}-w_{1}}{c^{2}}.
  \label{eq:redshift}
\end{equation}
Equations~(\ref{eq:gr-clock})--(\ref{eq:redshift}) are the standard 1PN formulae for the Global Navigation Satellite System (GNSS) (net $\sim$38~$\mu$s/day), for interplanetary cruise ($v\sim 30$~km\,s$^{-1}$ gives $1-\gamma^{-1}\sim 5\times 10^{-9}$), and for the annual modulation of ground-clock rates from the solar $w/c^{2}$ term along Earth's heliocentric orbit; Section~\ref{sec:iau} assembles these 1PN rates into the IAU TDB--TT chain.

\subsection{Christoffel Symbols for the 1PN Isotropic Potential}

For the static 1PN line element (\ref{eq:pn-metric}) with $w=w(\vec{x})$ and $\lambda\equiv w/c^{2}$, define $f=1-2\lambda$ and $h=1+2\lambda$. The non-vanishing Christoffel symbols to $\mathcal{O}(c^{-2})$ are \cite{will1993theory,damour2002coordinate}
\begin{align}
  \Gamma^{0}_{0i} &= -\frac{1}{c^{2}}\,\frac{\partial w}{\partial x^{i}}
  + \mathcal{O}(c^{-4}),
  \label{eq:gamma-00i}\\
  \Gamma^{i}_{00} &= -\frac{1}{c^{2}}\,\frac{\partial w}{\partial x^{i}}
  + \mathcal{O}(c^{-4}),
  \label{eq:gamma-i00}\\
  \Gamma^{i}_{jk} &= \frac{1}{c^{2}}\,\frac{\partial w}{\partial x^{i}}\,\delta_{jk}
  + \mathcal{O}(c^{-4}),
  \label{eq:gamma-ijk}
\end{align}
with all $\Gamma^{0}_{ij}=\Gamma^{i}_{0j}=0$ for this static, shift-free coordinate system. Timelike geodesics then yield the Newtonian limit $\mathrm{d}^{2}x^{i}/\mathrm{d}t^{2}=-\partial^{i}w$ and the redshift law (\ref{eq:gr-clock}); the $\Gamma^{i}_{jk}$ term drives the $\mathcal{O}(c^{-2})$ light deflection when applied to null curves.

\subsection{Christoffel Symbols in Harmonic PPN with $V_{i}$}

Retaining the gravitomagnetic potentials in (\ref{eq:ppn}), write the 1PN harmonic line element as
\begin{align}
  \mathrm{d}s^{2} &=
  \left(1-\frac{2w}{c^{2}}\right)c^{2}\mathrm{d}t^{2}
  + \frac{2W_{i}}{c^{2}}\,\mathrm{d}t\,\mathrm{d}x^{i}
  \notag\\
  &\quad - \left(1+\frac{2\gamma w}{c^{2}}\right)\mathrm{d}\vec{x}^{\,2},
  \label{eq:ppn-shift}\\
  W_{i} &\equiv (1+\gamma)V_{i},
  \notag
\end{align}
with $V_{i}$ the standard PPN vector potential \cite{will1993theory,soffel2003iau}. To $\mathcal{O}(c^{-3})$ the non-vanishing Christoffel symbols are
\begin{align}
  \Gamma^{0}_{00} &= 0,
  \label{eq:G000}\\
  \Gamma^{0}_{0i} &= -\frac{1}{c^{2}}\,\frac{\partial w}{\partial x^{i}},
  \label{eq:G00i}\\
  \Gamma^{0}_{ij} &= \frac{1}{c^{3}}\left(
  \frac{\partial W_{i}}{\partial x^{j}}
  + \frac{\partial W_{j}}{\partial x^{i}}\right),
  \label{eq:G0ij}\\
  \Gamma^{i}_{00} &=
  -\frac{1}{c^{2}}\,\frac{\partial w}{\partial x^{i}}
  + \frac{2}{c^{3}}\,\frac{\partial W_{i}}{\partial t},
  \label{eq:Gi00}\\
  \Gamma^{i}_{0j} &= \frac{1}{c^{2}}\left(
  \frac{\partial W_{i}}{\partial x^{j}}
  - \frac{\partial w}{\partial x^{i}}\,\delta_{ij}\right),
  \label{eq:Gi0j}\\
  \Gamma^{i}_{jk} &= \frac{\gamma}{c^{2}}\,
  \frac{\partial w}{\partial x^{i}}\,\delta_{jk},
  \label{eq:Gijk}
\end{align}
and $\Gamma^{0}_{ij}$ is symmetric in $(i,j)$. Equations~(\ref{eq:G00i})--(\ref{eq:Gijk}) reduce to (\ref{eq:gamma-00i})--(\ref{eq:gamma-ijk}) when $W_{i}=0$ and $\gamma=1$. Inserting (\ref{eq:G000})--(\ref{eq:Gijk}) into (\ref{eq:geodesic}) recovers the Einstein--Infeld--Hoffmann equations for massive bodies and the photon geodesic used in VLBI \cite{petit2010iers}.

\subsection{$\mathcal{O}(c^{-4})$ Christoffel Corrections}

Extending (\ref{eq:ppn-shift}) to $\mathcal{O}(c^{-4})$ with $\lambda\equiv w/c^{2}$,
\begin{align}
  \mathrm{d}s^{2} &=
  \left(1-2\lambda+2\beta\lambda^{2}\right)c^{2}\mathrm{d}t^{2}
  + \frac{2W_{i}}{c^{2}}\,\mathrm{d}t\,\mathrm{d}x^{i}
  \notag\\
  &\quad - \left[1+2\gamma\lambda+(4\gamma+2-2\beta)\lambda^{2}\right]
  \mathrm{d}\vec{x}^{\,2},
  \label{eq:ppn-2pn}
\end{align}
the inverse metric acquires $W^{2}/c^{4}$ and $\lambda^{2}$ terms \cite{will1993theory,damour2002coordinate}. The leading additions to (\ref{eq:G000})--(\ref{eq:Gijk}) are
\begin{align}
  \Gamma^{0}_{00} &= \frac{\beta-1}{c^{4}}\,\frac{\partial w}{\partial t}
  + \mathcal{O}(c^{-6}),
  \label{eq:G2-000}\\
  \Gamma^{0}_{0i} &= -\frac{1}{c^{2}}\,\frac{\partial w}{\partial x^{i}}
  + \frac{(\beta-1)\,w}{c^{4}}\,\frac{\partial w}{\partial x^{i}}
  + \mathcal{O}(c^{-6}),
  \label{eq:G2-00i}\\
  \Gamma^{i}_{00} &=
  -\frac{1}{c^{2}}\,\frac{\partial w}{\partial x^{i}}
  + \frac{2}{c^{3}}\,\frac{\partial W_{i}}{\partial t}
  \notag\\
  &\quad + \frac{2(\gamma-\beta)\,w}{c^{4}}\,
  \frac{\partial w}{\partial x^{i}}
  + \mathcal{O}(c^{-6}),
  \label{eq:G2-i00}\\
  \Gamma^{i}_{jk} &= \frac{\gamma}{c^{2}}\,
  \frac{\partial w}{\partial x^{i}}\,\delta_{jk}
  - \frac{2\gamma^{2}w}{c^{4}}\,
  \frac{\partial w}{\partial x^{i}}\,\delta_{jk}
  + \mathcal{O}(c^{-6}),
  \label{eq:G2-ijk}
\end{align}
while $\Gamma^{0}_{ij}$ and $\Gamma^{i}_{0j}$ pick up $\mathcal{O}(c^{-5})$ contributions from $\partial_{k}W_{l}\,\partial^{k}W^{l}$ and from the $\lambda^{2}$ part of $g_{ij}$. Equations~(\ref{eq:G2-00i})--(\ref{eq:G2-ijk}) modify the proper-time rate to
\begin{align}
  \frac{\mathrm{d}\tau}{\mathrm{d}t}
  &= 1+\frac{w}{c^{2}}-\frac{v^{2}}{2c^{2}}
  -\frac{w^{2}}{2c^{4}}
  \notag\\
  &\quad + \frac{(4\beta-3\gamma-3)\,w^{3}}{6c^{6}}
  + \mathcal{O}(c^{-8}),
  \label{eq:gr-clock-2pn}
\end{align}
which is the standard 2PN redshift formula in DE ephemerides \cite{petit2010iers}. At Jupiter-range accuracies the $w^{2}/c^{4}$ term reaches $\sim 10^{-15}$ and is usually retained; the $(\beta,\gamma)$-dependent cubic term is a parameterised test of GR.

\subsection{GCRS: Rotating Earth and Explicit $W_{i}$}

In the GCRS the scalar potential is $w=w_{\mathrm{E}}+w_{\mathrm{ext}}$ (Earth monopole plus external tidal potential), and the gravitomagnetic piece of a uniformly rotating body is \cite{soffel2003iau,petit2010iers}
\begin{equation}
  W_{i} =
  -\frac{G}{2c^{3}}\,
  \frac{\bigl(\vec{J}_{\oplus}\times\vec{r}\bigr)_{i}}{r^{3}}
  + \mathcal{O}(c^{-5}),
  \qquad
  r=|\vec{r}|,
  \label{eq:W-J}
\end{equation}
where $\vec{r}$ is the position vector from the geocenter and $\vec{J}_{\oplus}$ is Earth's spin angular momentum. A uniform-sphere estimate gives $\vec{J}_{\oplus}=\tfrac{2}{5}\,M_{\oplus}R_{\oplus}^{2}\,\vec{\omega}_{\oplus}\sim 7\times 10^{33}$~kg\,m$^{2}$\,s$^{-1}$; the value used below follows the polar moment of inertia $C_{\oplus}\approx 8.0\times 10^{37}$~kg\,m$^{2}$ with $\omega_{\oplus}=7.292\,115\times 10^{-5}$~rad\,s$^{-1}$, so $|\vec{J}_{\oplus}|=C_{\oplus}\omega_{\oplus}\approx 5.9\times 10^{33}$~kg\,m$^{2}$\,s$^{-1}$. The curl of (\ref{eq:W-J}) generates the Lense--Thirring precession through $\Gamma^{i}_{0j}$; at the equatorial surface ($R_{\oplus}=6.378\,137\times 10^{6}$~m) $|W|\sim GJ_{\oplus}/(2c^{3}R_{\oplus}^{2})\sim 2\times 10^{-16}$~m\,s$^{-1}$.

Substituting (\ref{eq:W-J}) into (\ref{eq:G0ij})--(\ref{eq:Gi0j}) gives the Sagnac contribution to one-way range rate. For a ground station at $\vec{r}_{G}$ and spacecraft at $\vec{r}_{S}$, with $\hat{n}=(\vec{r}_{S}-\vec{r}_{G})/\rho$,
\begin{align}
  \frac{1}{c^{2}}\,\frac{\mathrm{d}}{\mathrm{d}t}
  \bigl(W_{i}\,n^{i}\bigr)
  &\approx
  \frac{G}{2c^{5}\rho^{3}}\,
  \vec{n}\cdot\!\left[
  \bigl(\dot{\vec{r}}_{S}-\dot{\vec{r}}_{G}\bigr)\times\vec{J}_{\oplus}
  \right.
  \notag\\
  &\quad \left.
  - 3\bigl(\vec{n}\cdot\dot{\vec{r}}_{S}\bigr)\,
  \bigl(\vec{n}\times\vec{J}_{\oplus}\bigr)\times\vec{r}_{S}
  \right],
  \label{eq:sagnac-rate}
\end{align}
to leading order in $1/c$ \cite{ashby2003relativity,moyer2003formalism}. Equation~(\ref{eq:sagnac-rate}) enters the two-way Doppler law (\ref{eq:range-rate-1pn}) and is evaluated in TCG (or TT) when reducing ground-station data. The GCRS example fixes the Earth metric and Sagnac template used throughout; the IAU coordinate-time chain for Earth is developed in Section~\ref{sec:iau}, and the lunar and Mars body-centric extensions in Section~\ref{sec:body-crs}.

\section{IAU coordinate times and the transformation chain}
\label{sec:iau}

\subsection{Definitions}

The IAU 1991/2000 resolutions define TCG in the GCRS and TCB in the BCRS as coordinate times whose spatial origins coincide with the geocenter and the solar-system barycentre, respectively \cite{soffel2003iau,kaplan2005iau}. Terrestrial Time is a quasi-proper-time scale for geocentric work:
\begin{equation}
  \frac{\mathrm{d}t_{\mathrm{TT}}}{\mathrm{d}t_{\mathrm{TCG}}}
  = 1 - L_{\mathrm{G}}, \qquad
  L_{\mathrm{G}} = 6.969290134\times 10^{-10},
  \label{eq:tt-tcg}
\end{equation}
chosen so that the rate of TT matches that of an ideal clock on the geoid. TDB is a linear rescaling of TCB,
\begin{align}
  t_{\mathrm{TDB}} &= t_{\mathrm{TCB}}
  - L_{\mathrm{B}}\times(t_{\mathrm{TCB}}-t_{\mathrm{TCB},0})
  + \mathrm{periodic\ terms},
  \label{eq:tcb-tdb}\\
  L_{\mathrm{B}} &= 1.550519768\times 10^{-8},
  \notag
\end{align}
with periodic terms dominated by Earth's heliocentric motion ($\sim$1.7~ms peak-to-peak) so that, on average, $\langle t_{\mathrm{TDB}}-t_{\mathrm{TT}}\rangle$ has no secular drift at the geocenter \cite{fairhead1990analytical,fukushima1995time}. The solar $w/c^{2}$ modulation along Earth's BCRS orbit, together with the heliocentric velocity term in (\ref{eq:gr-clock}), is the dominant source of the periodic part of (\ref{eq:tcb-tdb}); analytical and numerical series are distributed in the IERS Conventions \cite{petit2010iers,brumberg1995essential}.

\subsection{TCB--TCG Relation}

The full 1PN transformation is an integral over the geocenter world line in the BCRS \cite{soffel2003iau}:
\begin{align}
  t_{\mathrm{TCG}} - t_{\mathrm{TCB}} &=
  \frac{L_{\mathrm{C}}}{c^{2}}\int\!\left(
  \frac{v_{\mathrm{E}}^{2}}{2} + w_{\mathrm{E}} + w_{\mathrm{E}}^{\mathrm{ext}}
  \right)\mathrm{d}t_{\mathrm{TCB}} \notag\\
  &\quad + \mathrm{periodic\ terms},
  \label{eq:tcb-tcg}
\end{align}
where $L_{\mathrm{C}} = 1.4808268677\times 10^{-8}$, $\vec{v}_{\mathrm{E}}$ is the geocenter velocity in the BCRS, $w_{\mathrm{E}}$ the Newtonian potential at the geocenter (including self-potential terms as specified in the resolution), and $w_{\mathrm{E}}^{\mathrm{ext}}$ the external potential from other bodies. Equation~(\ref{eq:tcb-tcg}) is the barycentric statement of (\ref{eq:gr-clock}) for the Earth world line. Substituting DE440-class ephemerides \cite{park2021de440,petit2010iers} yields the numerical series distributed in the IERS Conventions. The same 1PN prescription extends to other celestial bodies. Coordinate time is an integral of (\ref{eq:gr-clock}) along the parent-body centre-of-mass world line with the appropriate $w$ and $\vec{v}$. Section~\ref{sec:body-crs} develops the Mars (MCRS/MCG) case and the lunar (LCRS/TCL) analogue of the Earth chain (Figure~\ref{fig:timescales}). Consistency across CRS labels rests on documented transformations and radiometric closure, not on a global simultaneity convention \cite{damour2002coordinate}.

\section{Fermi normal coordinates and tetrad transport}
\label{sec:fermi}

Fermi normal coordinates supply the local tetrad frame for comparing distant clocks once the global metric of Sections~\ref{sec:kinematics}--\ref{sec:ppn} and the IAU coordinate times of Section~\ref{sec:iau} are fixed.

\subsection{Construction Along a Reference Geodesic}

Let $\gamma_{0}(\tau)$ be a timelike geodesic with four-velocity $u^{\alpha}$ and affine parameter $\tau$ (proper time). Fermi normal coordinates (FNC) $(\tau,\xi^{i})$ are defined by Synge's construction \cite{synge1960relativity}: at each event on $\gamma_{0}$, spacelike vectors $e^{\alpha}_{i}$ ($i=1,2,3$) form an orthonormal triad,
\begin{equation}
  g_{\alpha\beta}\,e^{\alpha}_{i}e^{\beta}_{j}=\delta_{ij},
  \qquad
  g_{\alpha\beta}\,u^{\alpha}e^{\beta}_{i}=0,
  \label{eq:tetrad-orthonormal}
\end{equation}
propagated along $\gamma_{0}$ by \emph{Fermi--Walker transport}. Using the covariant derivative (\ref{eq:covariant-dtau}), the transport law is
\begin{equation}
  \frac{\mathrm{D}e^{\alpha}_{i}}{\mathrm{D}\tau}
  = -\left(a_{\beta}u^{\alpha}\right)e^{\beta}_{i},
  \qquad
  a^{\alpha} \equiv \frac{\mathrm{D}u^{\alpha}}{\mathrm{D}\tau}
  = u^{\beta}\nabla_{\beta}u^{\alpha},
  \label{eq:fw}
\end{equation}
where $a^{\alpha}$ is the four-acceleration of $\gamma_{0}$. Equation~(\ref{eq:fw}) preserves both (\ref{eq:tetrad-orthonormal}) and orthogonality to $u^{\alpha}$; it differs from parallel transport ($\mathrm{D}e^{\alpha}_{i}/\mathrm{D}\tau=0$) whenever $a^{\alpha}\neq 0$, but coincides with parallel transport on a geodesic because then $a^{\alpha}=0$. For a non-geodesic world line (e.g.\ a spacecraft with thrust), the Fermi--Walker triad rotates with the four-acceleration through the term $a_{\beta}u^{\alpha}e^{\beta}_{i}$; inserting the tabulated $\Gamma^{\alpha}_{\beta\gamma}$ into (\ref{eq:covariant-dtau}) gives the explicit coordinate components. A neighbouring event is labelled by $(\tau,\xi^{i})$ through the exponential map; to quadratic order in $\xi$,
\begin{align}
  \mathrm{d}s^{2} &= -c^{2}\mathrm{d}\tau^{2}
  + \delta_{ij}\,\mathrm{d}\xi^{i}\mathrm{d}\xi^{j}
  - \frac{c^{2}}{3}R_{0i0j}\,\tau^{2}\,\mathrm{d}\xi^{i}\mathrm{d}\xi^{j} \notag\\
  &\quad - \frac{2c}{3}R_{0ij0}\,\tau\,\mathrm{d}\tau\,\mathrm{d}\xi^{i}
  + \mathcal{O}(\xi^{3}),
  \label{eq:fnc-metric}
\end{align}
with $R_{\mu\nu\rho\sigma}$ the Riemann tensor at $\gamma_{0}(\tau)$ \cite{synge1960relativity}. A clock at $\xi^{i}=0$ defines the FNC time $\tau$; a clock at fixed small $\xi^{i}$ experiences tidal redshift through $R_{0i0j}$. FNC are the standard local laboratory frame for stating that ``physics is locally special-relativistic'' without invoking a global inertial coordinate system.

\subsection{Remote Clock Comparison}

To compare proper times $\tau_{A}$ and $\tau_{B}$ at events $A$ and $B$, the comparison is defined by (i)~the coordinate system (BCRS, GCRS, or a body-centric CRS); (ii)~the world lines; (iii)~a \emph{time transformation}---either integrate (\ref{eq:gr-clock}) along a coordinate path joining $A$ and $B$, or transport a tetrad from $A$ to $B$ and evaluate the Doppler shift in that tetrad \cite{ashby2003relativity,damour2002coordinate}. The IAU resolutions of Section~\ref{sec:iau} implement the first route at 1PN accuracy with $w$ and $\vec{v}$ taken from a barycentric ephemeris. Along a timelike world line, the increment $\mathrm{d}\tau$ is generally different from the coordinate-time increment $\mathrm{d}t$ of the embedding CRS; at 1PN their ratio is given by (\ref{eq:gr-clock}), and in full generality by (\ref{eq:curved-tau}). Mapping on-board $\tau$ to TCB therefore involves the metric along the spacecraft arc, an explicit time transformation, and documented leap-second/TT handling at the ground segment.

\section{Null geodesics, Shapiro delay, and observables}
\label{sec:light}

Deep-space radiometry follows from null geodesics in the same 1PN metric. This section derives the light-time constraint, Shapiro delay, and the two-way Doppler law that closes the chain begun in Section~\ref{sec:kinematics}.

\subsection{Light-Time Equation}

Electromagnetic ranging treats the signal as a null geodesic ($\mathrm{d}s^{2}=0$). The retarded coordinate time of reception $t_{R}$ satisfies the light-time functional
\begin{equation}
  t_{R} = t_{T} + \frac{1}{c}\int_{\Gamma}\!
  \left(1+\frac{w}{c^{2}}+\cdots\right)\mathrm{d}\ell,
  \label{eq:light-time}
\end{equation}
where $t_{T}$ is transmission time, $\Gamma$ the ray path, $\mathrm{d}\ell$ the spatial line element along the ray, and $w$ the total scalar potential in the chosen CRS (including all gravitating bodies along $\Gamma$) \cite{moyer2003formalism,petit2010iers}. Partial derivatives $\partial\rho/\partial\vec{r}$ in orbit determination depend on whether $t_{T}$ is tagged in UTC, TT, or TDB.

\subsection{Shapiro Delay in PPN}

For a two-body configuration with mass $M$, transmitter at $\vec{r}_{1}$, receiver at $\vec{r}_{2}$, and ray passing at impact parameter $b$, the Shapiro correction to coordinate travel time is \cite{will1993theory,hellings1983relativistic}
\begin{equation}
  \Delta t_{\mathrm{Sh}}
  = -\frac{(1+\gamma)GM}{c^{3}}
  \ln\!\left(\frac{r_{1}+r_{2}+l}{r_{1}+r_{2}-l}\right)
  + \mathcal{O}(c^{-5}),
  \label{eq:shapiro}
\end{equation}
with $l=\sqrt{(r_{1}+r_{2})^{2}-b^{2}}$. On Earth--Mars links ($r_{1,2}\sim 1$--$1.5$~AU), omitting $\Delta t_{\mathrm{Sh}}$ biases the inferred range by hundreds of metres to kilometres through $c\,\Delta t_{\mathrm{Sh}}$, depending on the impact parameter $b$ (near-collinear rays with $b\to r_{1}+r_{2}$ are at the lower end); Sun-grazing geometries are larger still.

\subsection{Eikonal Equation and Derivation of the Delay}

For monochromatic radiation of coordinate angular frequency $\omega$, insert $\psi = A\,\mathrm{e}^{\mathrm{i}\omega(t-S/c)}$ into the curved-spacetime wave equation $\Box\psi=0$ and expand in $1/c$. The leading \emph{eikonal} equation is \cite{will1993theory,moyer2003formalism}
\begin{equation}
  g^{\mu\nu}\,\frac{\partial S}{\partial x^{\mu}}\,\frac{\partial S}{\partial x^{\nu}} = 0,
  \label{eq:eikonal}
\end{equation}
so the wave covector $k_{\mu}=\partial_{\mu}S$ is null. Along the ray, $k^{\mu}$ obeys the null geodesic equation $\mathrm{D}k^{\mu}/\mathrm{D}\lambda=0$ with affine parameter $\lambda$, where $\mathrm{D}/\mathrm{D}\lambda\equiv k^{\alpha}\nabla_{\alpha}$ is the covariant derivative along the ray (the null analogue of (\ref{eq:covariant-dtau})) \cite{will1993theory,moyer2003formalism}. For the 1PN metric (\ref{eq:pn-metric}), write $S=t+S_{1}+\mathcal{O}(c^{-2})$; substitution in (\ref{eq:eikonal}) gives $\|\vec{\nabla}S_{1}\|^{2}=0$ at $\mathcal{O}(c^{0})$ and, at $\mathcal{O}(c^{-2})$,
\begin{equation}
  \frac{\partial S_{1}}{\partial t}
  + \frac{2w}{c^{2}} + \|\vec{\nabla}S_{1}\|^{2} = 0.
  \label{eq:eikonal-1pn}
\end{equation}
Along a straight spatial path $\vec{x}(u)=\vec{x}_{T}+u\,\hat{n}$ ($0\le u\le\rho$) with $\hat{n}$ fixed at emission, $\vec{\nabla}S_{1}=\hat{n}$ and integration from transmission event $T$ to reception $R$ yields
\begin{align}
  t_{R}-t_{T}
  &= \frac{\rho}{c}\left(1+\frac{2\bar{w}}{c^{2}}\right)
  + \mathcal{O}(c^{-4}),
  \label{eq:eikonal-delay}\\
  \bar{w} &= \frac{1}{\rho}\int_{0}^{\rho}
  w\bigl(\vec{x}_{T}+u\hat{n}\bigr)\,\mathrm{d}u,
  \notag
\end{align}
the coordinate travel time to 1PN. Expanding $w$ about the ray midpoint and retaining the logarithmic term from the $1/r$ potential of a mass $M$ reproduces (\ref{eq:shapiro}) with factor $(1+\gamma)$ when $\gamma=1$ \cite{hellings1983relativistic}. Equation~(\ref{eq:eikonal-delay}) is the bridge between the null geodesic $\mathrm{d}s^{2}=0$ and the range integrals used in ephemeris software.

\subsection{Covariant DSN Observables and the Light-Time Constraint}

Deep Space Network tracking supplies two-way range and Doppler; their relativistic correction is a root-finding problem on the light-time equation subject to the null geodesic constraint \cite{moyer2003formalism,petit2010iers}. Let $x^{\mu}_{T}=(ct_{T},\vec{r}_{T})$ and $x^{\mu}_{R}=(ct_{R},\vec{r}_{R})$ be transmission and reception events in the BCRS, with $u^{\mu}_{T,R}$ the station and spacecraft four-velocities. A null signal satisfies
\begin{equation}
  g_{\mu\nu}\bigl(x^{\alpha}_{R}\bigr)\,k^{\mu}\,k^{\nu}=0,
  \qquad
  k^{\mu} = \frac{\mathrm{d}x^{\mu}}{\mathrm{d}\lambda},
  \label{eq:null-k}
\end{equation}
and the \emph{light-time constraint}
\begin{equation}
  \vec{r}_{R}(t_{R}) - \vec{r}_{T}(t_{T})
  = \rho\,\hat{n},
  \qquad
  \rho = c\bigl(\tau + \Delta t_{\mathrm{rel}}\bigr),
  \label{eq:ltc}
\end{equation}
where $\tau=t_{R}-t_{T}$ and $\Delta t_{\mathrm{rel}}$ collects (\ref{eq:eikonal-delay}) and (\ref{eq:shapiro}). Equation~(\ref{eq:ltc}) is solved iteratively because $t_{T}=t_{R}-\tau$ and $\vec{r}_{T}$ depend on the spacecraft ephemeris. The \emph{computed range} observable is $\rho=|\vec{r}_{R}-\vec{r}_{T}|$ at the retarded epoch.

The received-to-transmitted frequency ratio for a transponder or one-way carrier is the projection of $k^{\mu}$ on the receiver and transmitter four-velocities \cite{moyer2003formalism,ashby2003relativity}:
\begin{equation}
  \frac{\nu_{R}}{\nu_{T}}
  = \frac{\left(1+\lambda_{R}\right)\,k_{\mu}\,u^{\mu}_{R}}
         {\left(1+\lambda_{T}\right)\,k_{\mu}\,u^{\mu}_{T}}
  + \mathcal{O}(c^{-4}),
  \label{eq:doppler-cov}
\end{equation}
with $\lambda_{R,T}=w_{R,T}/c^{2}$ evaluated at the respective events and $k_{\mu}$ taken along the solved null ray.

\subsection{Two-Way Doppler: 1PN Expansion of $\dot{\rho}$}

This subsection expands the two-way geometric range rate $\dot{\rho}$, where $\rho$ is the instantaneous station--spacecraft range, to 1PN order.

\begin{figure}[t]
\centering
\includegraphics[width=\columnwidth]{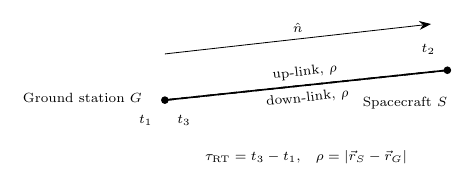}
\caption{Coherent Two-Way DSN Geometry ($t_{1}$: ground transmit; $t_{2}$: spacecraft turnaround; $t_{3}$: ground receive). Round-trip light time $\tau_{\mathrm{RT}}=t_{3}-t_{1}$ is obtained from the light-time constraint (Eq.~\ref{eq:ltc}) with 1PN and Shapiro corrections (Eqs.~\ref{eq:eikonal-delay}, \ref{eq:shapiro}).}
\label{fig:dsn-twoway}
\end{figure}

Coherent two-way DSN tracking measures the time derivative of the round-trip phase. Let the ground station transmit at $t_{1}$, the spacecraft receive and retransmit at $t_{2}$, and the station receive the echo at $t_{3}$. The \emph{round-trip light time} is
\begin{align}
  \tau_{\mathrm{RT}} &= (t_{3}-t_{1})
  = \frac{2\rho}{c}\left(1+\frac{2\bar{w}}{c^{2}}\right)
  \notag\\
  &\quad + \Delta t_{\mathrm{Sh}}^{(\mathrm{up})}
  + \Delta t_{\mathrm{Sh}}^{(\mathrm{dn})}
  + \mathcal{O}(c^{-4}),
  \label{eq:tau-rt}
\end{align}
with $\rho$ the instantaneous geometric range, $\bar{w}$ the mean potential along the ray from (\ref{eq:eikonal-delay}), and $\Delta t_{\mathrm{Sh}}^{(\mathrm{up,dn})}$ the Shapiro terms (\ref{eq:shapiro}) on each leg \cite{moyer2003formalism}. Differentiating (\ref{eq:tau-rt}) with respect to coordinate time $t$ at the station (TDB-labelled) gives the \emph{relativistic range rate}
\begin{align}
  \dot{\tau}_{\mathrm{RT}} &= \frac{2}{c}
  \left(1+\frac{2w}{c^{2}}\right)\dot{\rho}
  + \frac{2\rho}{c^{3}}\left(
  \frac{\partial w}{\partial t}
  + \vec{v}\cdot\vec{\nabla}w\right) \notag\\
  &\quad + 2\,\frac{\mathrm{d}\Delta t_{\mathrm{Sh}}}{\mathrm{d}t}
  + \frac{2}{c^{2}}\,\frac{\mathrm{d}}{\mathrm{d}t}
  \bigl(W_{i}\,n^{i}\bigr)
  + \mathcal{O}(c^{-4}),
  \label{eq:range-rate-1pn}
\end{align}
where $\vec{v}$ is the spacecraft velocity relative to the BCRS, $\hat{n}$ the unit vector along the ray, and $W_{i}=(1+\gamma)V_{i}$ from (\ref{eq:ppn-shift}). For a solar-mass monopole,
\begin{equation}
  \frac{\mathrm{d}\Delta t_{\mathrm{Sh}}}{\mathrm{d}t}
  = -\frac{(1+\gamma)GM_{\odot}}{c^{3}}
  \left(\frac{\dot{r}_{1}}{r_{1}}+\frac{\dot{r}_{2}}{r_{2}}\right)
  + \mathcal{O}(c^{-5}),
  \label{eq:shapiro-rate}
\end{equation}
with $r_{1,2}$ the heliocentric distances of station and spacecraft. The DSN \emph{two-way Doppler} observable is the fractional frequency shift
\begin{equation}
  \frac{\dot{f}_{D}}{f}
  = -\frac{\dot{\tau}_{\mathrm{RT}}}{\tau_{\mathrm{RT}}}
  = -\frac{c}{2\rho}\,\dot{\tau}_{\mathrm{RT}}
  + \mathcal{O}(c^{-4}),
  \label{eq:doppler-2way}
\end{equation}
for a coherent transponder that re-emits without remodulating the carrier \cite{moyer2003formalism,petit2010iers}. Expanding (\ref{eq:doppler-2way}) with (\ref{eq:range-rate-1pn}) yields the geometric term $-2\dot{\rho}/c$ multiplied by $(1+2w/c^{2})$ plus gravitational and Shapiro-rate corrections at the $10^{-12}$--$10^{-13}$ level for Mars-range arcs---comparable to carrier-tracking measurement uncertainty budgets. The one-way limit (\ref{eq:doppler-cov}) is recovered on a single leg by replacing $2\rho/c$ with $\rho/c$ and dropping the down-link Shapiro rate. Orbit-determination partials $\partial\dot{f}_{D}/\partial\vec{r}$ and $\partial\dot{f}_{D}/\partial\vec{v}$ chain through (\ref{eq:ltc}), (\ref{eq:range-rate-1pn}), and the TDB time tag of $t_{3}$.

\section{Body-centric celestial reference systems}
\label{sec:body-crs}

With the formalism of Sections~\ref{sec:kinematics}--\ref{sec:light} in place---in particular the IAU coordinate-time chain of Section~\ref{sec:iau}---this section applies the same 1PN template to bodies beyond Earth. It does \emph{not} re-derive the nested TCG--TCL series of Liu et al.\ \cite{liu2025convert}; that work supplies the leading lunar coefficients quoted below. Our role here is to embed LCRS/TCL and MCRS/MCG in the same metric--radiometry chain as Sections~\ref{sec:ppn}--\ref{sec:light} and to cross-check surface clock-rate estimates. Subsection~\ref{sec:mcrs-mars} calculates the Mars areoid--geoid instantaneous clock-rate difference (\ref{eq:mars-vs-earth}) from the MCRS metric, the provisional MCG coordinate time (\ref{eq:mars-tcb})--(\ref{eq:mars-mcg-periodic}), and the surface comparison to a geoid-referenced TT clock. No literature leading-rate coefficient analogous to (\ref{eq:tcl-tcg}) is yet available. Subsection~\ref{sec:lcrs-validation} repeats the chain for LCRS and TCL and compares the monopole and nested TCG--TCL rates with \cite{liu2025convert,kopeikin2024lunar,ashby2024relativistic}.

\subsection{MCRS: Mars gravity field, MCG prescription, and areoid--geoid clock rate}
\label{sec:mcrs-mars}

The MCRS is an areocentric celestial reference system with origin at the Mars system barycentre (for precision work, the areocentre) and spatial axes tied to the IAU Mars orientation model \cite{kaplan2005iau,soffel2003iau}. Unlike the Moon, where nested TCG--TCL leading coefficients are taken from clock-comparison literature, the areoid--geoid offset (\ref{eq:mars-vs-earth}) follows from the MCRS metric, the provisional MCG prescription, and the 1PN redshift chain. MCG remains a provisional symbol pending IAU fixing of $L_{\mathrm{M}}$.

MCRS is embedded in the BCRS through the Mars ephemeris $\vec{r}_{\mathrm{M}}(t)$ and velocity $\vec{v}_{\mathrm{M}}(t)$ from DE440-class integrations \cite{park2021de440}. Denote by $\vec{r}'$ the position relative to the areocentre. The 1PN line element is (\ref{eq:ppn-shift}) with
\begin{align}
  w &= \frac{GM_{\mathrm{M}}}{r'}
  + w_{\mathrm{ext}}(\vec{r}'),
  \label{eq:mars-W-J}\\
  W_{i} &=
  -\frac{G}{2c^{3}}\,
  \frac{\bigl(\vec{J}_{\mathrm{M}}\times\vec{r}'\bigr)_{i}}{r'^{\,3}}
  + \mathcal{O}(c^{-5}),
  \notag
\end{align}
where $r'=|\vec{r}'|$, $w_{\mathrm{ext}}$ collects the external tidal potential from the Sun and planets exterior to Mars (chiefly the Sun and Jupiter), and $\vec{J}_{\mathrm{M}}$ is the Mars spin angular momentum. For a uniform-sphere model,
\begin{align}
  \vec{J}_{\mathrm{M}}
  &= \frac{2}{5}\,M_{\mathrm{M}}\,R_{\mathrm{M}}^{2}\,\vec{\omega}_{\mathrm{M}},
  \label{eq:mars-J}\\
  \omega_{\mathrm{M}} &= \frac{2\pi}{T_{\mathrm{M}}},
  \notag
\end{align}
with $M_{\mathrm{M}}=6.417\,1\times 10^{23}$~kg, $R_{\mathrm{M}}=3.396\,2\times 10^{6}$~m, and Mars sidereal rotation period $T_{\mathrm{M}}=88\,775.2$~s ($\omega_{\mathrm{M}}=7.088\,1\times 10^{-5}$~rad\,s$^{-1}$), giving $|\vec{J}_{\mathrm{M}}|\approx 2.1\times 10^{32}$~kg\,m$^{2}$\,s$^{-1}$. At the equatorial surface, $|W|\sim 2\times 10^{-17}$~m\,s$^{-1}$---far below DSN thresholds but retained in the metric for consistency with (\ref{eq:G0ij})--(\ref{eq:Gi0j}).

\begin{figure}[t]
\centering
\includegraphics[width=\columnwidth]{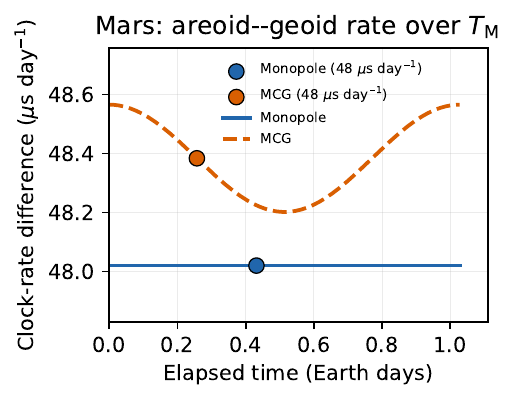}
\caption{Mars areoid--geoid instantaneous clock-rate difference over $T_{\mathrm{M}}$ (schematic). Solid curve: monopole rate from (\ref{eq:mars-vs-earth}); dashed curve: (\ref{eq:mars-rate-schematic}).}
\label{fig:clock-drift-mars}
\end{figure}

MCG is introduced by the same 1PN prescription as TCG: coordinate time in the MCRS whose rate differs from TCB by integration of (\ref{eq:gr-clock}) along the areocentre world line in the BCRS,
\begin{equation}
  t_{\mathrm{MCG}} - t_{\mathrm{TCB}}
  = \frac{L_{\mathrm{M}}}{c^{2}}\int\!
  \mathcal{H}_{\mathrm{M}}\,\mathrm{d}t_{\mathrm{TCB}}
  + \mathrm{periodic\ terms},
  \label{eq:mars-tcb}
\end{equation}
with $\mathcal{H}_{\mathrm{M}}=v_{\mathrm{M}}^{2}/2+w_{\mathrm{M}}+w_{\mathrm{M}}^{\mathrm{ext}}$ and $L_{\mathrm{M}}$ a dimensionless constant fixed by a chosen Mars quasi-proper-time scale (analogous to $L_{\mathrm{G}}$ for TT). By the same 1PN logic as (\ref{eq:moon-tcl-periodic}), the periodic part of (\ref{eq:mars-tcb}) is a Fourier spectrum at harmonics of the areocentre heliocentric motion and Mars spin,
\begin{equation}
  \frac{\mathrm{d}t_{\mathrm{MCG}}}{\mathrm{d}t_{\mathrm{TCB}}}
  - \left\langle\frac{\mathrm{d}t_{\mathrm{MCG}}}{\mathrm{d}t_{\mathrm{TCB}}}\right\rangle
  = \sum_{k=1}^{\infty}\bigl(C_{k}\cos k\,\lambda_{\mathrm{M}}
  + D_{k}\sin k\,\lambda_{\mathrm{M}}\bigr),
  \label{eq:mars-mcg-periodic}
\end{equation}
with $\lambda_{\mathrm{M}}$ a mean-anomaly-like argument from Mars ephemeris and $(C_k,D_k)$ from a DE440-class series \cite{park2021de440,petit2010iers}. The corresponding areoid--geoid instantaneous rate inherits this periodic content on top of the secular monopole value (\ref{eq:mars-vs-earth}).
A clock on the areoid at rest in the Mars-fixed frame satisfies, from (\ref{eq:gr-clock}) and (\ref{eq:gr-clock-2pn}),
\begin{equation}
  \frac{\mathrm{d}\tau}{\mathrm{d}t_{\mathrm{MCG}}}
  = 1 - \frac{GM_{\mathrm{M}}}{c^{2}R_{\mathrm{M}}}
  + \frac{\omega_{\mathrm{M}}^{2}R_{\mathrm{M}}^{2}}{2c^{2}}
  + \mathcal{O}(c^{-4}),
  \label{eq:mars-clock}
\end{equation}
where the centrifugal term uses the equatorial radius. Relative to a geoid-referenced TT clock,
\begin{align}
  \frac{\mathrm{d}\tau_{\mathrm{M}}}{\mathrm{d}\tau_{\oplus}}
  &\approx 1 + \frac{\Delta w_{\mathrm{M}\oplus}}{c^{2}}
  \notag\\
  &\approx 1 + 5.6\times 10^{-10},
  \label{eq:mars-vs-earth}
\end{align}
with $\Delta w_{\mathrm{M}\oplus}=w_{\mathrm{M}}-w_{\oplus}
\approx GM_{\oplus}/R_{\oplus}-GM_{\mathrm{M}}/R_{\mathrm{M}}$
at leading order (surface potentials $w\simeq -GM/R$; $R_{\oplus}=6.371\times 10^{6}$~m),
the monopole magnitude difference at rest on the two surfaces (tides and centrifugal corrections enter at higher order).
Inserting $M_{\mathrm{M}}=6.417\,1\times10^{23}$~kg and $R_{\mathrm{M}}=3.396\,2\times10^{6}$~m gives the monopole estimate $\Delta w_{\mathrm{M}\oplus}/c^{2}=5.558\times10^{-10}$, i.e.\ $48.0$~$\mu$s\,day$^{-1}$. The centrifugal correction in (\ref{eq:mars-clock}) contributes only $\sim3.2\times10^{-13}$ to $\mathrm{d}\tau/\mathrm{d}t_{\mathrm{MCG}}-1$ at the equator. External $w_{\mathrm{M}}^{\mathrm{ext}}$ enters at higher order in the surface comparison. Both terms are retained in the metric for consistency but do not alter the quoted monopole rate at the precision targeted here. This purely metric offset is distinguished from oscillator drift in \textbf{Tianwen-1} lander and relay timing.
Figure~\ref{fig:clock-drift-mars} compares the monopole rate (\ref{eq:mars-vs-earth}) with a single-harmonic schematic correction over one sidereal day $T_{\mathrm{M}}$,
\begin{equation}
  \frac{\mathrm{d}\tau_{\mathrm{M}}}{\mathrm{d}\tau_{\oplus}} - 1
  \approx 5.6\times 10^{-10}
  + A_{\mathrm{M}}\cos\!\left(\frac{2\pi t}{T_{\mathrm{M}}}\right),
  \label{eq:mars-rate-schematic}
\end{equation}
with $A_{\mathrm{M}}\approx2.1\times10^{-12}$; a precision model would use the full sum (\ref{eq:mars-mcg-periodic}) with DE440-class $(C_k,D_k)$.
For a spacecraft in low Mars orbit at altitude $h$, replace $R_{\mathrm{M}}$ by $R_{\mathrm{M}}+h$ in (\ref{eq:mars-clock}) and add $\vec{v}_{\mathrm{orb}}\cdot\hat{n}$ through (\ref{eq:range-rate-1pn}); substituting (\ref{eq:mars-W-J}) into (\ref{eq:sagnac-rate}) with $\vec{J}_{\mathrm{M}}$ and MCRS velocities yields the Mars Sagnac correction on two-way links between Zhurong, the Tianwen-1 orbiter, and Earth stations.

\begin{figure*}[t]
\centering
\includegraphics[width=\textwidth]{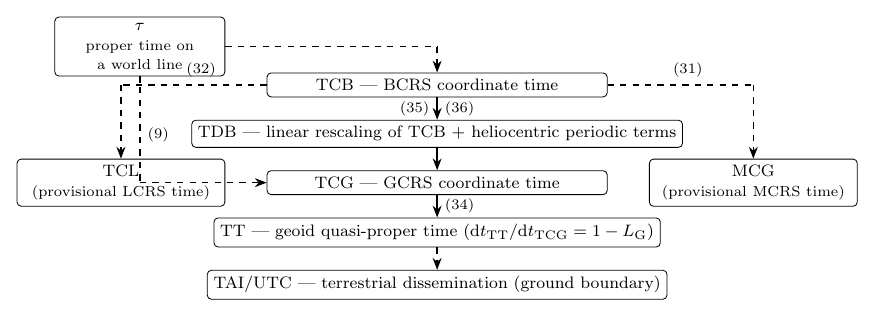}
\caption{Solar-System Time-Scale Hierarchy. Solid arrows: IAU coordinate-time transformations; dashed links: proper-time mapping, time dissemination (TT to TAI/UTC), and the provisional LCRS/MCRS extensions (TCL, MCG). Labels (9), (31)--(32), (34)--(36) refer to Equations~(\ref{eq:gr-clock}), (\ref{eq:mars-tcb}), (\ref{eq:moon-tcb}), (\ref{eq:tt-tcg}), (\ref{eq:tcb-tdb}), and (\ref{eq:tcb-tcg}).}
\label{fig:timescales}
\end{figure*}
Figure~\ref{fig:timescales} collects the hierarchy of Table~\ref{tab:timescales} together with the provisional LCRS and MCRS extensions.

\subsection{LCRS: Lunar gravity field and $\vec{J}_{\mathrm{L}}$}
\label{sec:lcrs-validation}

The lunar case applies the prescription of (\ref{eq:tcb-tcg}) and (\ref{eq:tt-tcg}) with selenocentre quantities, using the same body-centric chain as in Subsection~\ref{sec:mcrs-mars}. The LCRS is a selenocentric celestial reference system with origin at the Moon's centre of mass (selenocentre) and spatial axes tied to the IAU lunar orientation model \cite{kaplan2005iau,soffel2003iau,liu2025convert}. It is embedded in the BCRS through the lunar ephemeris $\vec{r}_{\mathrm{L}}(t)$ and velocity $\vec{v}_{\mathrm{L}}(t)$ from DE440-class integrations \cite{park2021de440}. Denote by $\vec{r}'$ the position relative to the selenocentre. The 1PN line element is (\ref{eq:ppn-shift}) with
\begin{align}
  w &= \frac{GM_{\mathrm{L}}}{r'}
  + w_{\mathrm{ext}}(\vec{r}'),
  \label{eq:moon-W-J}\\
  W_{i} &=
  -\frac{G}{2c^{3}}\,
  \frac{\bigl(\vec{J}_{\mathrm{L}}\times\vec{r}'\bigr)_{i}}{r'^{\,3}}
  + \mathcal{O}(c^{-5}),
  \notag
\end{align}
where $r'=|\vec{r}'|$, $w_{\mathrm{ext}}$ collects the external tidal potential from the Earth, Sun, and other bodies exterior to the Moon, and $\vec{J}_{\mathrm{L}}$ is the lunar spin angular momentum. For a uniform-sphere model of the synchronously rotating Moon,
\begin{align}
  \vec{J}_{\mathrm{L}}
  &= \frac{2}{5}\,M_{\mathrm{L}}\,R_{\mathrm{L}}^{2}\,\vec{\omega}_{\mathrm{L}},
  \label{eq:moon-J}\\
  \omega_{\mathrm{L}} &= \frac{2\pi}{T_{\mathrm{L}}},
  \notag
\end{align}
with $M_{\mathrm{L}}=7.342\times 10^{22}$~kg, $R_{\mathrm{L}}=1.737\,4\times 10^{6}$~m, and sidereal spin period $T_{\mathrm{L}}=2.360\,591\,5\times 10^{6}$~s ($\omega_{\mathrm{L}}=2.661\,7\times 10^{-6}$~rad\,s$^{-1}$), equal to the sidereal orbital period under synchronous rotation. At the equatorial surface, $|W|\sim 10^{-18}$~m\,s$^{-1}$---well below DSN and lunar laser ranging (LLR) thresholds but retained for consistency with (\ref{eq:G0ij})--(\ref{eq:Gi0j}).

\begin{figure}[t]
\centering
\includegraphics[width=\columnwidth]{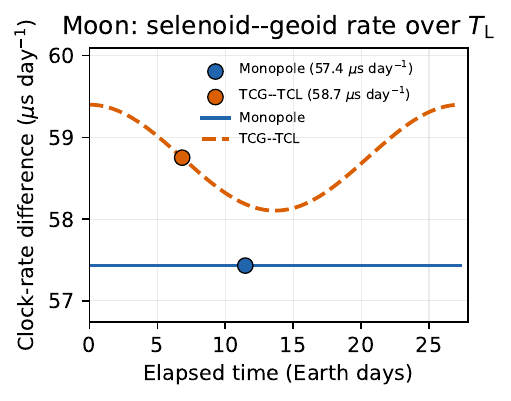}
\caption{Lunar selenoid--geoid instantaneous clock-rate difference over $T_{\mathrm{L}}$ (schematic). Solid curve: monopole rate from (\ref{eq:moon-vs-earth}); dashed curve: (\ref{eq:moon-rate-schematic}). Markers at $\sim$57.4 and $\sim$58.7~$\mu$s\,day$^{-1}$ follow \cite{liu2025convert,kopeikin2024lunar,ashby2024relativistic}.}
\label{fig:clock-drift-moon}
\end{figure}

\emph{Lunar Coordinate Time} (TCL; symbol provisional) is introduced by the same 1PN prescription as TCG: coordinate time in the LCRS whose rate differs from TCB by integration of (\ref{eq:gr-clock}) along the selenocentre world line in the BCRS,
\begin{equation}
  t_{\mathrm{TCL}} - t_{\mathrm{TCB}}
  = \frac{L_{\mathrm{L}}}{c^{2}}\int\!
  \mathcal{H}_{\mathrm{L}}\,\mathrm{d}t_{\mathrm{TCB}}
  + \mathrm{periodic\ terms},
  \label{eq:moon-tcb}
\end{equation}
with $\mathcal{H}_{\mathrm{L}}=v_{\mathrm{L}}^{2}/2+w_{\mathrm{L}}+w_{\mathrm{L}}^{\mathrm{ext}}$ and $L_{\mathrm{L}}$ a dimensionless constant fixed by a chosen lunar quasi-proper-time scale (analogous to $L_{\mathrm{G}}$ for TT). Nested Earth--Moon reference systems give a direct TCG--TCL relation whose leading rate coefficient can be refined from clock comparisons \cite{liu2025convert,kopeikin2024lunar,ashby2024relativistic}:
\begin{equation}
  \frac{\mathrm{d}t_{\mathrm{TCL}}}{\mathrm{d}t_{\mathrm{TCG}}}
  \approx 1 + 6.8\times 10^{-10}
  + \mathrm{periodic\ terms}.
  \label{eq:tcl-tcg}
\end{equation}
At 1PN order the ``$+$ periodic terms'' in (\ref{eq:moon-tcb}) and (\ref{eq:tcl-tcg}) are the non-secular remainder of integrating (\ref{eq:gr-clock}) along the selenocentre world line in the BCRS. Because $\mathcal{H}_{\mathrm{L}}(t)=v_{\mathrm{L}}^{2}(t)/2+w_{\mathrm{L}}(t)+w_{\mathrm{L}}^{\mathrm{ext}}(t)$ varies with lunar orbital phase---through DE440-class $\vec{r}_{\mathrm{L}}(t)$, $\vec{v}_{\mathrm{L}}(t)$ and the Earth's tidal potential at the Moon---the TCG--TCL rate acquires a Fourier spectrum at harmonics of the lunar mean motion $n_{\mathrm{L}}=2\pi/T_{\mathrm{L}}$:
\begin{equation}
  \frac{\mathrm{d}t_{\mathrm{TCL}}}{\mathrm{d}t_{\mathrm{TCG}}}
  - 1 - 6.8\times 10^{-10}
  = \sum_{k=1}^{\infty}\bigl(A_{k}\cos k\,\lambda_{\mathrm{L}}
  + B_{k}\sin k\,\lambda_{\mathrm{L}}\bigr),
  \label{eq:moon-tcl-periodic}
\end{equation}
with $\lambda_{\mathrm{L}}=n_{\mathrm{L}}(t_{\mathrm{TCG}}-t_0)$ and coefficients $(A_k,B_k)$ supplied by nested Earth--Moon time-transformation series \cite{liu2025convert,kopeikin2024lunar,petit2010iers}. Differentiating (\ref{eq:moon-tcb}) makes the same periodic content appear as an \emph{instantaneous} rate modulation on top of the secular mean in (\ref{eq:tcl-tcg}), distinct from the constant monopole estimate (\ref{eq:moon-vs-earth}).
LTC denotes the proposed operational scale. It is the lunar counterpart of UTC, realised by atomic clocks on or near the Moon, steered to TCL, and disseminated to users \cite{liu2025rule,liu2025convert,whitehouse2024celestial}. A clock on the selenoid at rest in the Moon-fixed frame satisfies, from (\ref{eq:gr-clock}) and (\ref{eq:gr-clock-2pn}),
\begin{equation}
  \frac{\mathrm{d}\tau}{\mathrm{d}t_{\mathrm{TCL}}}
  = 1 - \frac{GM_{\mathrm{L}}}{c^{2}R_{\mathrm{L}}}
  + \frac{\omega_{\mathrm{L}}^{2}R_{\mathrm{L}}^{2}}{2c^{2}}
  + \mathcal{O}(c^{-4}),
  \label{eq:moon-clock}
\end{equation}
where the centrifugal term is negligible for the slow lunar spin. Relative to a geoid-referenced TT clock,
\begin{align}
  \frac{\mathrm{d}\tau_{\mathrm{L}}}{\mathrm{d}\tau_{\oplus}}
  &\approx 1 + \frac{\Delta w_{\mathrm{L}\oplus}}{c^{2}}
  \approx 1 + 6.65\times 10^{-10},
  \label{eq:moon-vs-earth}
\end{align}
with $\Delta w_{\mathrm{L}\oplus}=w_{\mathrm{L}}-w_{\oplus}
\approx GM_{\oplus}/R_{\oplus}-GM_{\mathrm{L}}/R_{\mathrm{L}}$
at leading order (surface potentials $w\simeq -GM/R$; tides and centrifugal corrections enter at higher order).
Inserting $M_{\oplus}=5.972\times 10^{24}$~kg, $R_{\oplus}=6.371\times 10^{6}$~m, and the lunar values above gives $\Delta w_{\mathrm{L}\oplus}/c^{2}\approx 6.65\times 10^{-10}$, i.e.\ $6.65\times10^{-10}\times86\,400\times10^{6}\approx57.4$~$\mu$s\,day$^{-1}$ from the monopole surface comparison (\ref{eq:moon-vs-earth}).
Equation~(\ref{eq:tcl-tcg}) quotes the nested TCG--TCL leading rate $6.8\times 10^{-10}$ from clock-comparison work \cite{liu2025convert,kopeikin2024lunar,ashby2024relativistic}, equivalent to $\sim$58.7~$\mu$s\,day$^{-1}$. The $\sim$2\% spread between (\ref{eq:moon-vs-earth}) and (\ref{eq:tcl-tcg}) reflects selenocentre orbital velocity, the external Earth-tide potential $w_{\mathrm{L}}^{\mathrm{ext}}$, and geoid--selenoid structure beyond the $GM/R$ surface comparison.
Figure~\ref{fig:clock-drift-moon} compares the monopole rate (\ref{eq:moon-vs-earth}) with a schematic modulation anchored to (\ref{eq:tcl-tcg}) over one sidereal month $T_{\mathrm{L}}$,
\begin{equation}
  \frac{\mathrm{d}\tau_{\mathrm{L}}}{\mathrm{d}\tau_{\oplus}} - 1
  \approx 6.8\times 10^{-10}
  + A_{\mathrm{L}}\cos\!\left(\frac{2\pi t}{T_{\mathrm{L}}}\right),
  \label{eq:moon-rate-schematic}
\end{equation}
with $A_{\mathrm{L}}=\tfrac{1}{2}(6.8-6.65)\times10^{-10}\approx7.5\times10^{-12}$; a precision model would use the full sum (\ref{eq:moon-tcl-periodic}) with ephemeris $(A_k,B_k)$ \cite{liu2025convert}.
This metric drift is distinguished from oscillator drift in \textbf{Chang'e-4/6} far-side relay timing, cislunar navigation, and LLR \cite{liu2025convert,liu2025rule}. For a spacecraft in low lunar orbit at altitude $h$, replace $R_{\mathrm{L}}$ by $R_{\mathrm{L}}+h$ in (\ref{eq:moon-clock}) and add $\vec{v}_{\mathrm{orb}}\cdot\hat{n}$ through (\ref{eq:range-rate-1pn}); substituting (\ref{eq:moon-W-J}) into (\ref{eq:sagnac-rate}) with $\vec{J}_{\oplus}$ and BCRS/GCRS velocities yields the Earth Sagnac correction on two-way Earth--Moon links.

\section{Operational implementation}
\label{sec:operations}

Section~\ref{sec:body-crs} quotes body-centric clock rates; this section states how the documented transformation chains enter merged tracking, standards, and orbit-determination pipelines.

\subsection{Radiometry and Mission Clocks}

Round-trip light time $\Delta t=t_{R}-t_{T}$ corrected by (\ref{eq:shapiro}) and plasma terms yields range $\rho\approx c\Delta t/2$ at the few-metre level when timing is controlled to nanoseconds. \textbf{Chang'e} cislunar missions and \textbf{Tianwen-1} (Mars system, 2020--) illustrate the time-scale specification issue. Ground hydrogen masers are steered to UTC/TAI. Spacecraft oscillators measure $\tau$ along cislunar or heliocentric world lines. Science products carry documentation of whether event times are UTC, mission elapsed time (MET), TDB, or a provisional TCL/LTC label. They also record the transformation between proper time $\tau$ and $t_{\mathrm{TCB}}$ or between $\tau$ and $t_{\mathrm{TCL}}$ used in orbit determination \cite{moyer2003formalism,petit2010iers,liu2025rule}. Merging domestic and foreign tracking without that metadata introduces microsecond-level systematic offsets in range and Doppler. On Earth--Mars links, the uncompensated areoid--geoid drift is $\sim$48~$\mu$s\,day$^{-1}$ from the theoretical monopole rate (\ref{eq:mars-vs-earth}). On Earth--Moon links, the selenoid--geoid drift spans the literature monopole secular rate $\sim$57.4~$\mu$s\,day$^{-1}$ (\ref{eq:moon-vs-earth}) and the nested TCG--TCL leading rate $\sim$58.7~$\mu$s\,day$^{-1}$ (\ref{eq:tcl-tcg}).

\subsection{Standards, Transfer, and Software}

Caesium fountains and optical lattice clocks realise the SI second on Earth \cite{guinot2005time}; flight rubidium/hydrogen standards act as transfer oscillators whose relation to TAI is established by two-way links during passes. Orbit-determination pipelines implement (\ref{eq:tt-tcg})--(\ref{eq:shapiro}), leap-second tables, and station-position models from the IERS Conventions under the TCB/TDB tagging conventions of planetary ephemerides \cite{standish1998time,petit2010iers}. Relativistic residuals in lunar laser ranging, planetary laser ranging, and radio science provide independent checks of $(\gamma,\beta)$ and of the TCB implementation.

\section{Conclusions}
\label{sec:conclusion}

We presented a unified 1PN documentation chain that links the PPN harmonic metric, IAU coordinate times, body-centric CRS extensions, and two-way DSN observables within established resolutions. Tabulated Christoffel symbols through $\mathcal{O}(c^{-4})$ connect the line element to the redshift and geodesic routines used in IERS-class software. The Earth-centred chain (\ref{eq:tt-tcg})--(\ref{eq:tcb-tcg}) extends to MCRS/MCG and LCRS/TCL in Section~\ref{sec:body-crs}. Fermi normal coordinates (\ref{eq:fnc-metric}) specify local clock comparison. Null geodesics yield the light-time constraint (\ref{eq:ltc}), Shapiro delay (\ref{eq:shapiro}), and the 1PN range-rate expansion (\ref{eq:range-rate-1pn})--(\ref{eq:doppler-2way}) used in coherent tracking (Figure~\ref{fig:dsn-twoway}).

For Mars, the MCRS/MCG prescription gives an areoid--geoid instantaneous clock-rate offset of $5.558\times10^{-10}$ ($\sim$48.0~$\mu$s\,day$^{-1}$) from (\ref{eq:mars-W-J})--(\ref{eq:mars-vs-earth}). The periodic band is much smaller (\ref{eq:mars-rate-schematic}, \ref{eq:mars-mcg-periodic}). Figure~\ref{fig:clock-drift-mars} shows the rate over $T_{\mathrm{M}}$. On the Moon, the monopole rate $\sim$57.4~$\mu$s\,day$^{-1}$ (\ref{eq:moon-vs-earth}) and nested leading rate $\sim$58.7~$\mu$s\,day$^{-1}$ (\ref{eq:tcl-tcg}) agree with published clock-transform coefficients \cite{liu2025convert,kopeikin2024lunar,ashby2024relativistic}. Figure~\ref{fig:clock-drift-moon} displays the larger lunar periodic modulation (\ref{eq:moon-rate-schematic}, \ref{eq:moon-tcl-periodic}). The 1PN range-rate law (\ref{eq:range-rate-1pn}) exposes gravitational and Shapiro-rate terms at $10^{-12}$--$10^{-13}$ on Mars-range arcs. These terms are comparable to current carrier-phase noise floors.

For Chang'e- and Tianwen-class operations, merged radiometry can acquire microsecond-level biases unless event times carry explicit TDB tags, on-board proper-time maps, and leap-second metadata. The documented sequence from proper time $\tau$ to $t_{\mathrm{TCL}}$, then to $t_{\mathrm{TCB}}$ and $t_{\mathrm{TT}}$ is needed on lunar missions. On Mars missions, the corresponding sequence runs from $\tau$ to $t_{\mathrm{MCG}}$, then to $t_{\mathrm{TCB}}$ and $t_{\mathrm{TT}}$. Remaining tasks include $\mathcal{O}(c^{-6})$ ephemeris terms, IAU-fixed constants $L_{\mathrm{L}}$ and $L_{\mathrm{M}}$, an operational LTC realisation, outer-planet CRS extensions, and joint tests with optical clocks and multi-agency links.

\section*{Acknowledgements}
This work is supported by the Strategic Priority Research
  Program of the Chinese Academy of Sciences (grant No. XDA0350304), the National Astronomical Observatories of the Chinese Academy of Sciences (Project No.~E4TG6601), and the National Key Research and Development Program of China (Grant No.~2021YFC2203000).

\bibliographystyle{plainnat}
\bibliography{ref}

\end{document}